# Strategic Insights from Simulation Gaming of AI Race Dynamics

## Authors

Ross Gruetzemacher, Shahar Avin, James Fox, Alexander K Saeri

## Abstract

We present insights from `Intelligence Rising', a scenario exploration exercise about possible AI futures. Drawing on the experiences of facilitators who have overseen 43 games over a four-year period, we illuminate recurring patterns, strategies, and decision-making processes observed during gameplay. Our analysis reveals key strategic considerations about AI development trajectories in this simulated environment, including: the destabilising effects of AI races, the crucial role of international cooperation in mitigating catastrophic risks, the challenges of aligning corporate and national interests, and the potential for rapid, transformative change in AI capabilities. We highlight places where we believe the game has been effective in exposing participants to the complexities and uncertainties inherent in AI governance. Key recurring gameplay themes include the emergence of international agreements, challenges to the robustness of such agreements, the critical role of cybersecurity in AI development, and the potential for unexpected crises to dramatically alter AI trajectories. By documenting these insights, we aim to provide valuable foresight for policymakers, industry leaders, and researchers navigating the complex landscape of AI development and governance.

## Contact

Ross Gruetzemacher (ross.gruetzemacher@wichita.edu)

## Affiliations

| | |
|---|---|
| Ross Gruetzemacher | Barton School of Business, Wichita State University<br>Transformative Futures Institute<br>Centre for the Study of Existential Risk, University of Cambridge |
| Shahar Avin | Centre for the Study of Existential Risk, University of Cambridge<br>Centre for the Future of Intelligence, University of Cambridge |
| James Fox | Department of Computer Science, Oxford University<br>London Initiative for Safe AI (LISA) |
| Alexander K Saeri | BehaviourWorks Australia, Monash Sustainable Development Institute, Monash University |

# Table of Contents

# Executive Summary

Intelligence Rising is a tabletop exercise that allows participants to simulate AI race dynamics between states (i.e., the U.S. and China) and AI developers (e.g., Alphabet, Microsoft, Baidu) as they all seek to develop radically transformative AI. Intelligence Rising originated in 2018 as a free form scenario exercise and was developed into a more rigid tabletop exercise in 2019.

In this manuscript the three lead facilitators of Intelligence Rising detail the lessons and insights that we gleaned over the course of nearly four years of playing Intelligence Rising — from September 2020 to July 2024. A structured elicitation was conducted for the facilitators by an independent party in order to help answer the three research questions that were chosen as the focus on this reflection on the games played. We briefly outline these in the Findings subsection below.

## FIndings

**Research Question 1: What key insights about the role and impact of AI over the next 10 years need to be communicated to decision-makers and policymakers?**

1. Current and future AI systems pose a wide variety of societal-scale risks, ranging from imminent threats like disinformation and unfair outcomes to longer-term catastrophic and existential risks.
2. Even prior to the development of radically transformative AI, AI technologies can have dramatically destabilising effects as they rapidly advance and reshape society.
3. The power to steer the future of AI development is very unequally distributed due to several drivers for concentration, including the enormous compute requirements of the latest frontier AI models.
4. There exists an information asymmetry where states and the public will constantly be catching up to deal with the impacts of the last generation of AI technologies, unless they manage to regulate the rate of progress or invest in anticipating new developments and impacts.
5. Cooperation is a major challenge.

**Research question 2: What are some recurring storylines, strategies, or player behaviours that have emerged from the game?**

We identified twelve salient clusters that emerged during our elicitation regarding recurring storylines, strategies, and players' actions:

1. *Espionage and cyberwarfare:* cyberwarfare is generally widespread, especially for espionage.
2. *Winners take all rather than winner takes all:* teams often reach radically transformative AI simultaneously leading to multipolar outcomes.
3. *Tech race dynamics:* a race dynamic generally emerges between tech firms, with firms emphasising safety to governments but often deprioritizing it internally.
4. *Races are destabilising:* races often result in blocs split by national allegiance where the "loser" uses military actions to prevent the "winner" deploying radically transformative AI.

5. *Division into blocks by state lines:* blocs typically form based on national allegiance, with U.S. tech firms aligning with the U.S. and Chinese tech firms aligning with China.
6. *Taiwan often comes up but is not determinative:* hard power is sometimes used before the endgame, mostly over Taiwan, which can impact on semiconductor supply.
7. *Cooperation is hard, usually only achieved through top down intervention (e.g., PPP or nationalisation):* true nationalisation is easier in China whereas public private partnerships (PPPs) are easier in the U.S., and even when agreements are in place to slow progress and focus on safety ensuring trust is difficult and deception is common.
8. *Government policy reactive and slow; instability tends to increase:* policy development is slow, typically more reactive than proactive, and the slow progress can be destabilising.
9. *Elections are disruptive to US long-term strategic cohesion, important for catastrophic risk:* U.S. politics and elections are burdensome on leaders with elections resulting in a transfer of executive power to the right can negatively impact AI policy efforts; China is generally more consistent and rational with respect to efforts to mitigate risk.
10. *Autonomous weapons systems:* There is often tension between U.S. tech firms and U.S. leaders on the use of AI in autonomous weapons but agreements are made over policy or resources.
11. *Players try to accumulate resources:* all players try to accumulate the critical resources powering AI progress like AI researchers, computational resources, and data.
12. *Supply chain disruptions slow but don't stop AI and cause instability:* disruptions to the semiconductor supply chain slow but do not stop progress and are destabilising.

**Research Question 3: How have facilitators' perspectives or insights on strategic AI futures evolved through conducting this game?**

***Geopolitical Race for AI***

- As a result of our experiences facilitating Intelligence Rising, our concerns over the potential consequences of a race, particularly the potential for a direct military conflict spurred by concerns over AI, have grown significantly.
- We have grown more accepting of the notion of China 'winning' an AI race over time, especially when an agreement is in place to ensure the autonomy of Western states.
- We have come to see an agreement assuring the autonomy of both democratic and non-democratic governments as being critical to success.

***Technical and Policy Solutions***

- We have come to see that agreements ensuring states' autonomy and values between states of opposing ideologies are often a necessary condition for optimal outcomes when RTAI is created.
- We have also come to realise that there are many other significant challenges to crafting agreements of this nature, and work on better understanding the nature of such agreements is an understudied topic in the realm of AI strategy research.
- We have come to realise that in reality it is not likely that labs and states will all contribute equally to safety, yet, despite this, we now realise the critical role that coordination on safety research may play.

***Roleplay***

- It has become apparent that even U.S. citizens are not adept in quickly switching from portraying a U.S. president of one U.S. party (e.g., Democrat) to portraying a U.S. president of the opposite U.S. party (e.g., Republican).
- During this period we have seen an increase in the prior knowledge of our participants related to AI risk and safety issues as well as an increase in our players' abilities to confidently portray different actors in the different scenarios in the game.

***Uncertainty***

- We have realised during our time facilitating the game that there is a broad range of perceptions as to how progress toward advanced AI might transpire.
- This has taught us to respect the need for policymakers and decision-makers in AI firms to be very adaptable in the AI strategy space and to try to come up with governance proposals that are versatile to a wide range of possible events, trajectories and discontinuities.
- Our experiences from facilitating games has only made the inherent uncertainty of this complex problem more visceral, and we have come to oscillate between pessimism and optimism over whether the outcome will be favourable for humanity.
- We all expect the world 10 or 20 years from now to be dramatically different from the world today.

***Rate of Change***

- Nonetheless, even having experienced the rapid progress of the last several years, it is often still difficult to accept the implications of progress continuing at a similar or accelerated pace, as modelled in our game.
- This has, to varying degrees, afforded us all with more visceral experiences of the rapid pace of progress, and can sometimes pose challenges for us to harmonise our experiences with those of our daily lives.
- While we were still unprepared for our system one (Kahneman 2011) reactions to experiencing such tremendous progress deployed in a product with little warning, we were still able to use our system two processing to reorientate our models of the state of progress and pivot our work accordingly.

***Actors Modelled in the Game***

- Affecting meaningful change in how AI is developed and governed requires more than just providing information or engaging in lobbying efforts.
- Identifying and leveraging the key influences on these powerful entities is perhaps the most crucial aspect of working towards responsible AI development.

***Information Asymmetry***

- It is unclear what is actually going on behind the scenes, and those who only have access to public information are likely to have a limited or flawed perception of the true nature of the situation.

***State Actors***

- One of the most common things seen in all games was the exfiltration of AI model weights or the theft of algorithmic secrets (e.g., see RAND's 2024 report on Securing AI Model Weights).

## Conclusions

Here we highlight some of our conclusions:

- The unprecedented pace of technological progress in foundation models (Bommasani et al. 2021) presents novel challenges that make it very difficult for experts and non-experts alike to develop a bigger picture perspective on this progress.
- Outcomes leading to positive futures almost always require coordination between actors who by default have strong incentives to compete — this applies both to companies and to nations.
- Actors with sufficient resources and understanding of the risks are rare, and have an outsized potential to increase risk (by racing ahead, by developing destabilising applications such as autonomous weapons and cyberweapons, or by open sourcing advanced technologies) or to increase safety (by contributing significantly to the technical AI safety "commons" or by playing a coordination role amongst industry and state actors).
- Technology development does not happen in isolation — it affects, and is affected by, geopolitics, economical factors, social factors, and state actions. Actors should consider the broader consequences of their policies, including on trust between powerful actors, and impacts on social stability. There is no predetermined path that AI technology is bound to follow.

# Introduction

As AI has seen tremendous progress in the past five years (OpenAI 2023a; Wei et al. 2022; Bommasani et al. 2021; Brown et al. 2020), and AI safety has become a widely discussed topic over the past two years (Bengio et al. 2024; Cohen et al. 2024; Hendrycks et al. 2023; Critch and Russell 2023) since the public release of ChatGPT (OpenAI 2022) in November of 2022. In these discussions, there has been significant concern over catastrophic and existential risks (e.g., societal-scale risks; Hendrycks et al. 2023; Critch and Russell 2023; Gruetzemacher et al. 2024; Grace et al. 2024) associated with advanced AI such as artificial general intelligence (AGI; Goertzel and Wang 2007) or superintelligence (Bostrom 2014). However, concern over such societal-scale AI risks from advanced AI is not new (Cave et al. 2020; Good 1966).

The subject of this paper is Intelligence Rising (Avin et al. 2020), a strategic simulation game intended to let participants explore the space of plausible AI futures. The purpose of Intelligence Rising is two-fold: to explore the space of plausible AI futures and to raise awareness of the catastrophic and existential risks associated with advanced AI. The game specifically deals with conditions resembling a race to advanced AI among four or more major stakeholders (national governments or tech companies; e.g., China, the United States, Google, Microsoft, Baidu, or Tencent). Intelligence Rising allows participants to experience scenarios that decision-makers may encounter in coming years related to AI research and development (R&D), AI governance and geopolitics, and societal-scale risks from AI.

Intelligence Rising was initially developed in 2018. Through the first half of 2019, the game was a free-form role-play game and completely unstructured. The first complete structured version of the game was developed over a five-day design sprint workshop at the Future of Humanity Institute (FHI) in Oxford, UK in late summer of 2019 (Sandberg 2024). A second design sprint workshop was held at FHI in February of 2020 to further develop the game. Work pivoted to the development of an online version of the game later in the spring of 2020, and an initial online version of Intelligence Rising began testing in April of 2020. Over the course of the next year and a half, further development of the online game proceeded, and in-person games resumed in 2022.

Previous published work on Intelligence Rising described an initial version of the game (Avin et al. 2020), and forthcoming work will discuss evaluations that have been conducted with the games' participants to better understand the impact of the game (Mani et al. forthcoming). This paper examines another dimension of the game: the lessons learned by the facilitators who have facilitated the game over the last four years.

This paper is intended to summarise lessons from almost 200 hours of facilitated gameplay over 43 games. We hope these will be useful for those new to the field of AI risk or working in fields where the information could be helpful to decision-makers developing strategic plans or policy initiatives related to societal-scale risks from AI. Games have been found to be one of the most feasible techniques for anticipating AI futures (Gruetzemacher et al. 2021), and this is the first study to report results utilising this approach. To structure our reflection, we selected three specific research questions to focus on during our analysis of our experiences:

- What key insights about the role and impact of AI over the next 10 years need to be communicated to decision-makers and policymakers?
- What are some recurring storylines, strategies, or player behaviours that have emerged from the game?
- How have facilitators' perspectives or insights on strategic AI futures evolved through conducting this game?

To address these questions, a structured elicitation, led by an elicitation expert, was conducted for Intelligence Rising's three senior facilitators. This elicitation covered lessons learnt from 43 games facilitated after the second design sprint conducted at FHI in February 2020, with games being conducted between September 2020 and July 2024.

The remainder of this paper proceeds by first discussing extant literature related to wargaming and scenario exploration methods. This is followed by a description of Intelligence Rising in its current form, which was largely consistent over the period when the relevant games were being facilitated. Subsequently, the results from the elicitation of facilitator reflections are presented and discussed, leading to recommendations for both research and practice. In conclusion, we summarise the key insights.

# Literature Review

Catastrophic and existential risks from advanced AI have long been a topic of both academic (Bostrom and Crikovic 2011; Good 1966) and public interest (e.g., the Terminator movies; see Garvey 2018). With recent progress in AI, since late 2022 (OpenAI 2022; OpenAI 2023a), interest in catastrophic and existential risk from advanced AI has increased, from both academics (Bengio et al. 2024; Cohen et al. 2024; Critch and Russell 2023; Hendrycks et al. 2023) and the government (Frontier AI Taskforce 2023; EO 14110).

Work on the use of foresight approaches for exploring futures involving catastrophic or existential risks from advanced AI is limited (Baum et al. 2020; Beard et al. 2020; Avin 2019; Gruetzemacher 2019). Wargaming has long been used for futures exploration and for training military officers (Perla 1990), having been used as early as in the 1960s for gaming nuclear weapons catastrophes (Dalkey 1967) and trying to minimise casualties from hypothetical nuclear conflict (Owen 1969). Today, advanced approaches are used heavily in national security and defence communities for policy analysis (Bartels 2020), and there are efforts to develop next generation wargames (Reddie et al. 2018), the most recent of which are beginning to employ AI technologies similar to some of those explored in Intelligence Rising (Jensen et al. 2024).

Recently, futures methods have garnered interest with respect to AI (Korinek 2023), albeit more scenario-based approaches rather than gaming. Scenario-based approaches are useful for strategic planning in that they allow decision-makers to plan and prepare for various scenarios that are plausible but not necessarily likely. Scenario analysis has been in use since at least the 1960s when it was pioneered by leading strategic thinkers such as Herman Kahn at RAND (Kahn and Wiener 1967; Huss 1988), although the first academic paper on a comprehensive method for scenario creation dates to Zentner in 1975. Perhaps the most famous case of scenario analysis was by the Royal Dutch Shell Corporation during the 1973 oil crisis (Cornelius et al. 2005), in which Shell effectively utilised scenario planning to significantly outperform others in the sector in navigating the fallout from this shock event. This case study was ultimately effective at popularising the technique in the business world (Wack 1985a; 1985b). The Shell scenario planning approach is commonly associated with the intuitive logics school of scenario planning, which was derived from the work of Herman Kahn and is the most widely discussed technique in academic literature (Amer et al. 2013).

While Perla's (1990) The Art of Wargaming is quintessential reading for anyone new to wargaming, the domain still lacks an authoritative or comprehensive review of the range of techniques most often used, especially for the broad scope of applications wherein the techniques are used, ranging from military applications, to policy applications, to strategic planning applications in organisations. Unlike scenario planning techniques, much about wargames is siloed away in the organisations most familiar with their use like military universities or think tanks. Often, these different organisations will have distinct approaches to wargaming that reflect the best practices of the collective experience of their lead game designers. In the subsequent paragraphs, we try to synthesise wargaming literature from various perspectives, each of which in some way is relevant to Intelligence Rising.

While opinions vary about the value and best uses of wargaming (e.g., see Curry 2020; Perla and McGrady 2011; Perla 1985), it is important to note that wargaming should not necessarily be limited to military applications. Business wargames, or strategy wargames (West et al. 2018), are known to have applications in strategic planning across a wide variety of business domains (Chusil 2007) and have become more widely used in the past two decades (Gilad 2008). However, similar to the lack of a comprehensive source on wargaming methodology, and less than desired academic wargaming literature more broadly, there is relatively scant academic work on the use of wargaming in organisations, something that might be expected to be better documented in academic journals. There appears to be a much greater ratio of content published in practitioner journals to content published in academic journals than in other disciplines (e.g., see Courtney et al. 2020; Nichols 2013; Koller et al. 2012; Horn 2011).[1] Moreover, both practitioners and researchers propose wargame methodologies as a form of scenario planning (Cares and Miskel 2007; Schwartz et al. 2019). Wargames are particularly well-suited for some business applications, including response to supply chain disruptions (Milne and Longworth 2020) and cybersecurity incident response planning (Bailey et al. 2013).[2] It is unclear why relatively little academic work exists on wargames, but it may be related to the fact that the benefits are greater for larger organisations and research opportunities are limited for most academics without access to such organisations.[3]

While there is some literature on business wargames, in the design of Intelligence Rising we more closely reviewed the body of literature associated with more traditional wargames. As already described, wargaming approaches vary from organisation to organisation, or, from application to application. Policy wargames fall into neither business wargames nor more traditional wargames, yet are still an especially effective application of the methodology which has been significantly refined by practitioners. One of the most comprehensive reviews of wargames in the context of supporting policy design and policy-making was presented by Bartels (2020), who describes four archetypes of games that can be used to inform policy: system exploration, alternative conditions, innovation, and evaluation. System exploration games elicit experts' opinions on their mental models of a policy problem and how it may evolve over time. Alternative conditions games seek to explore similarities and differences in games' decision-making environments to understand the influence of varying conditions on decision-making. Innovation games are intended to identify new decision options that break away from the status quo. Evaluation games seek to judge the actions of participants based on a normative standard.

All of the above types of games described by Bartels (2020) involve data collection, which is necessary for informing policy or supporting decision-making. However, games can be used for a variety of pedagogical purposes as well (De Freitas 2018; Gros 2007; Kirriermuir and McFarlane 2004). Moreover, games are thought to be particularly well-suited for experiential learning (Gouveia et al. 2011; Saenz and Cano 2004). Intelligence Rising is more closely aligned with pedagogical games utilising experiential learning, at least in its objective; however, it is still relatively unique in its implementation for several reasons.

Returning back to more traditional wargaming, there are three fundamental types of games: decision games (a.k.a. tactical decision games), tabletop exercises (TTXs), and multi-day games or full-scale wargames. The first of these is the shortest and simplest option to implement, whereas TTXs take additional time to play and prepare, and multi-day games require the most time to prepare for and to play. The granularity of the information that one can obtain increases with the length of time spent planning and playing the game; however, games that generate the most granular information are not always best suited for all scenarios. Rather, a cost-benefit analysis is required to determine the optimal balance between planning time, the fidelity of the simulated world, and the granularity of the data to collect.

Decision games are excellent for quick game development and are useful for both analytical and educational purposes (Perla 1990). They often involve providing participants with a prompt and then asking participants to respond to a series of questions based on the prompt. The facilitator will then ask questions of the participants regarding their responses to the questions — i.e., the decisions they have made — and their reasoning for their responses. Their origin dates to their use by the British during WWII (Haustrath 1971), and they require the least amount of time to develop of the three types of games discussed here, being able to be run in one to two hours, and being able to be run in a virtual environment without significant loss in the quality of the data collected. Consequently, they are especially useful in collecting data and sometimes for scoping the breadth of scenarios to explore more granularly using TTXs or multi-day games.

---

[1] Note that the majority of these citations are linked to references printed in McKinsey Quarterly, with just one being from Harvard Business Review, potentially further underscoring their use in practice.

[2] Both of these topics often arise in games of Intelligence Rising, though the strategic focus of the game does not afford a detailed exploration of these topics when they arise during games.

[3] This would support the higher prevalence of wargaming being discussed in publications produced by firms like McKinsey, with access to large clients. Another consideration could be that McKinsey is more open to using the "wargaming" language, while other practical business publications are not, but this still does nothing to explain the relative paucity of academic work when compared with the abundance of practical work on the topic.

TTXs such as simulation games and serious games have long been used for pedagogical applications, particularly in the humanities (Aldrich 2005; Dorn 1989; Connolly 2013; Bellotti et al. 2011; Clark 2007). Aldrich (2005) was early to suggest their potential value in pedagogical information technology (IT) applications. In keeping with these projections, TTXs have recently been seeing successful applications in such pedagogical IT applications such as the adoption of disruptive technologies (de Rosa and Strode 2022) and cybersecurity training (Angafor 2020). More often than not, pedagogical simulation games and serious games are TTXs rather than decision games or full-scale wargames.

There is often no need for adjudication of actions taken by players in decision games, as the decisions are only used for discussion and have no consequences or bearing on the unfolding of the scenario. Even if adjudication is a component of decision games, the outcomes are generally predetermined, as with a Choose Your Own Adventure story. However, there are different forms of adjudication that can be utilised in the other forms of wargaming, e.g., TTXs and full-scale/multi-day wargames. In some cases, the means of adjudication is a central element of the game style, such as in matrix games where matrices are used for adjudicating claims as in decision-making applications outside of gaming (Herrero 2010). The matrices can be used to structure quantitative evaluation of actions based on resources possessed and/or allocated for the actions being resolved. Sometimes white cells can be used for adjudication of game actions when objectivity is needed to maintain fidelity or to prevent scepticism — either voiced or unspoken — of the game facilitator(s). In the context, a white cell refers to one or more persons participating in the game in the explicit role of adjudicator.

TTX adjudication is often simple such that a single adjudicator or facilitator is able to manage the entire game. In the most simple of cases this might involve adjudication directly by the facilitator. However, TTX games are also often simplified versions of complex scenarios that would require full-scale wargames to simulate with high fidelity. In such cases, when modelling more complex behaviour requires some structured rules for adjudication, matrices — i.e., matrix games — are most often utilised.

Finally, we consider full-scale wargames. Full-scale wargames are the most complex type of game and often involve multiple days, and sometimes even longer than a week to play through. They are used both for training purposes and data collection purposes. They are the highest quality games for modelling very complex interactions and system behaviour; however, the increased quality associated with them comes at a steep cost in terms of time and resources. Not only do these games take a substantial amount of experts' time over the days or weeks they take to play, but they also take anywhere from months up to a year or longer to plan. Further, they frequently required numerous white cell participants, facilitators, and note-takers, underscoring the significant burden they can be for designers and facilitators.

Intelligence Rising, the game being discussed in this article, is a TTX primarily designed for pedagogical purposes. However, due to its nature and broad, unbounded action space, it is nonetheless useful for exploring plausible futures. The following subsection describes Intelligence Rising in some further detail.

# Methods

This methods section describes the Intelligence Rising game and its evolution over time, the facilitators, the participants, and the process by which the facilitators' reflections were collected and analysed. A discussion of the facilitators and an analysis of the 43 games considered in this study are also included.

## The "Intelligence Rising" Game

An initial version of an AI strategy roleplay exercise was first explored by Owen Cotton-Barratt and Michael Page, then at the Future of Humanity Institute (FHI) in Oxford, and Shahar Avin at the Centre for the Study of Existential Risk (CSER) in Cambridge in July 2017. The unstructured exercise was iterated on by Shahar Avin, James Fox and Ross Gruetzemacher at CSER over the next two years. In September of 2019, with a broader set of collaborators,[4] a design sprint was conducted at FHI where the initial freeform version of the AI strategy roleplay game was expanded into the first, tabletop version of "Intelligence Rising", described in Avin, Gruetzemacher and Fox (2020). Key features of this game included:

- The division of players into four teams, two representing states and two representing multinational technology corporations.
- A turn-based decision-making loop where all teams commit to actions without knowledge of other team's actions for that turn, followed by simultaneous resolution by the facilitator.
- The per-turn decision-making was divided into two components: the selection of open-ended "policy actions" and an "AI R&D allocation." Limiting the "policy actions" to two per turn introduced an attention economy.
- A predefined "technology tree" that outlines dependencies in AI R&D, including basic research and possible applications, that the players explore via AI R&D allocation.
- A numerical tracking of four resources, including three abstract "powers" — "soft power", "hard power", and "cyber power" — as well as an abstract "budget", all of which are influenced by, and used to resolve, the freeform "policy actions".
- Teams were given the option to keep either policy actions or AI R&D allocation secret from other teams, which would then be resolved privately by the facilitator.

Following a second design iteration (based on experience with running the first tabletop version) and a third design iteration (based on the need to adapt to online play during the COVID-19 pandemic), a new version of the game was created, which we refer to here as "Intelligence Rising Online". This new version drew on a wider pool of expertise, including in risk communication and game design.[5] Key differences that were introduced in these iterations include:

- Switching to online platforms for gameplay (mainly Google Slides and Discord).
- Formalisation of team victory conditions and collective loss conditions.
- Simplification of the technology tree into two "lanes", each advancing through increased "levels". One lane represents future advances in language and world-modelling AI capabilities, and the other represents advances in AI reinforcement learning in increasingly complex environments.
- Addition of predefined "concerns" — negative consequences of AI applications that can potentially be mitigated through player actions.

The game format still required significant on-the-spot judgement from the facilitator, when resolving freeform policy actions and narrating the events of the game. There were also small variations between how different facilitators facilitated the game, for example when deciding how to resolve actions such as the nationalisation of a corporation, or when deciding on how to resolve the degree of safety of an advanced AI system. Lessons from the different experiences were fed back to the rest of the team, allowing facilitators to learn from each other and for small modifications to be made to the game assets and rules (though no significant changes were made to the core rules and assets of "Intelligence Rising

[4] Including Eran Dror, Anders Sandberg, James Fox, Ross Gruetzemacher and Shahar Avin.

[5] Contributors include all of those listed in the previous footnote, as well as Lara Mani, Markus Salmela, Eran Aviram, and Craig Chirouaki-Lewin.

Online" over the period reported in this paper). We acknowledge the limitations of this reflection paper, both from the variations between games facilitated by the different facilitators, which makes their experiences less comparable and by the regular exchange between facilitators, which makes their reflections not independent of each other.

As we continued facilitating "Intelligence Rising Online" after the pandemic, eventually we began to play in-person games again. Thus, some of the games reported here were conducted in person, with players using laptops or mobile phones to access the digital game assets during gameplay. We do not separate online and in-person games in the reflections below. We did notice differences between online and in-person games, but they did not directly impact the narratives that emerged. Some differences that we observed included:

- Participants were often more engaged during in-person games (e.g. less distracted by emails)
- Participants tended to be more creative in in-person games (often in-person debate amongst teammates fostered more creative actions, that were more challenging to recreate in video chat rooms)
- Facilitators found it easier to track a comprehensive picture of the world's state and the overarching narrative of games during in-person games; this was more difficult in online games, with many parallel video chats and private messages passed among players.

A significant design change was introduced to Intelligence Rising in 2023 through a novel game mechanic of a "Policy Tree", reflecting advances and consolidation of scholarly research on policies for governing frontier AI. Games that included the policy tree mechanic are excluded from the analysis reported in this paper.

# Facilitators

The reflections in this paper are drawn from the experiences of three Intelligence Rising facilitators, who all co-created the game and are co-authors of this paper. Given the nature of the game as a facilitated roleplay exercise, and in particular, given the importance of facilitator resolution of freeform policy actions (which form the bulk of decisions made in the game), the scenarios that unfold in Intelligence Rising rely on the background knowledge of the facilitators. The subject matter domain of Intelligence Rising includes highly advanced and speculative future technical AI research, the productization and business strategy underlying the progress of a very powerful general purpose technology (Eloundou et al. 2024; Gruetzemacher and Whittlestone 2022), and the core of international relations between great powers. Given the breadth and depth of the domain, and the lack of constraints on the players' action space, adjudication of actions in real-time can be challenging. We therefore find it relevant to summarise the backgrounds and experience of the facilitators in Table 1.

**Table 1:** Intelligence Rising Facilitators

| Name | Relevant Education | Relevant Experience | First Game | Number of games facilitated 09/2020-07/2024[6] |
|---|---|---|---|---|
| **Shahar Avin** | PhD in History and Philosophy of Science | Senior Research Associate at the Centre for the Study of Existential Risk and the Centre for the Future of Intelligence, University of Cambridge | 2017 | 20 |

[6] Some games were facilitated by more than one facilitator, hence the sum of this column is greater than the number of games covered by this paper (43).

| | | | | |
|---|---|---|---|---|
| **James Fox** | DPhil in Computer Science, University of Oxford | Research Director of the London Initiative for Safe AI (LISA), Visiting Researcher at the University of Oxford | 2020 | 19 |
| **Ross Gruetzemacher** | PhD in Management Information Systems, Dissertation: *Forecasting Transformative AI* | Assistant Professor of Business Analytics at Wichita State University; Executive Director of the Transformative Futures Institute; Research Affiliate at the Centre for the Study of Existential Risk | 2019 | 15 |

There are obvious limitations, both to the game and the analysis presented here, given that personal and professional biases are introduced at several stages:

- Recruitment of new facilitators is carried out by previous facilitators, through professional networks, limiting diversity.
- The number of facilitators to date has been small.[7]
- The facilitators both helped design the game and are the ones facilitating it, thus both explicitly and implicitly pushing narratives in particular directions.
- The facilitators are also authors of this paper.

While we do think the findings from the facilitators' experiences reported in this article will prove useful to readers, we also feel it is important for readers to keep these limitations in mind.

# Participants

Over the period from September 2020 to July 2024, the three facilitators identified in Table 1 facilitated 43 games of Intelligence Rising. Of these, 5 took place at conferences, where participants did not know each other prior to the game, and were unpaid. In the remaining 38 games, a pre-existing group arranged for its members to play Intelligence Rising, of which 22 were paid workshops and 17 were pro-bono; the breakdown of games by group type is as follows: AI company (4), insurance company (2), think tank or NGO (7), university course (7), non-academic training course (6), academic group (3), student groups (6), other (3). Some details of these games are listed in Appendix A.

The number of participants in conference games[8] ranged from 20 to 60, and in non-conference games from 4 to 20, usually with 8-12 participants, with the "team" size most frequently being 2-3 participants. While generally participants did have some knowledge of the AI domain, their backgrounds with respect to their prior experience with AI development and/or AI strategy varied widely — from lay knowledge to domain experts — mostly between games, but also within games.

The creation of the online version allowed us to facilitate Intelligence Rising remotely, but the recruitment of participant groups through word-of-mouth meant they were still relatively socially clustered (around groups interested in AI strategy and AI risks) and relatively geographically clustered (21 UK, 7 USA, 3 Belgium, 2 Sweden, 1 France, 1 Singapore, 8 fully remote).

[7] This was the case during the period of the games facilitated and analysed for this study. However, the number of facilitators on the team has been increasing since 2023, and we hope to share lessons from additional facilitators in future publications.

[8] All conference games were played at conferences of the Effective Altruism community. This population is generally very interested in and aware of AI risks, and many in the population work directly on the issue.

In most games covered by this paper, the participant cohort was more knowledgeable about AI technologies and AI impacts than the general population. In at least 10 games, more than half of the participants were actively working in the field of AI development or AI governance. In an additional 15 games, more than half of the participants were in the process of training to join the fields of AI development or AI governance. Finally, in an additional 14 games, at least a couple of the participants were either actively working in these fields or training to join them. In total, at least 39 of 43 games included at least some participants with expert knowledge of the domain. This feeds into the realism of the actions and narratives generated in the games.

## Data collection procedure, materials, and measures

While acknowledging the biases inherent in a reflections paper (outlined above in the Facilitators section), the authors adopted a structured process for collating and aggregating reflections.

To assist with recall (and also to support ongoing game development and facilitator training), after the completion of each game, facilitators were encouraged to fill out a game documentation form. This form covered logistics details of the game (date and location), the participant group, any modifications made to the game format before or during gameplay, and a narrative recounting of the key events of the game in the early, mid, late, and end stages of the game. The form also asks facilitators to rank the presence of eleven key themes in the specific playthrough, including agreements between states or companies, military actions or threats, espionage and cyberoperations, and competitions for R&D talent. In the relevant time period (09/2020-07/2024) there were 17 responses to the form, covering 40% of games (17/43). While all facilitators had access to the form reports of other facilitators, it was rare for these to be accessed. In preparing for the study reported in this paper, a non-facilitator author (Alexander Saeri) prepared refresher documents for each of the three facilitators with the form responses of the game they had run, which did not contain the form responses for games facilitated by other facilitators. In addition to the game documentation form, for about a quarter of games the facilitator for that game also prepared a report for the client, which covered similar topics as the game documentation form; facilitators were also encouraged to revisit those in preparation for the elicitation exercise.

In a discussion led by Alexander Saeri, the three facilitators agreed on three key questions to explore in this reflection study. These three research questions are:

1. What key insights about the role and impact of AI over the next 10 years need to be communicated to decision-makers and policymakers?
2. What are some recurring storylines, strategies, or player behaviours that have emerged from the game?
3. How have facilitators' perspectives or insights on strategic AI futures evolved through conducting this game?

Then, guided by the three questions and assisted by the refresher documents based on the game documentation form responses, each facilitator independently collected a list of answers, in bullet point format, for each of the questions.

Finally, a structured aggregation session led by Alexander Saeri was conducted using Miro. Individual answers to each of the questions were presented as virtual post-it notes, and then collectively clustered by the three Intelligence Rising facilitators. The resulting clusters were then labelled during discussion, and further observations about each cluster were generated through discussion. These aggregated answers to each of the three research questions are presented and discussed in the next section.

# Results and Discussion

This section is organised by the three research questions that are the focus of this analysis. To report the key insights, it made more sense to discuss each insight in prose. For the clusters of recurring game elements, we use a table to organise the data and subsequently discuss these results. Finally, for reporting the reflections on salient topics, we again use a table to organise our collective thoughts and follow this with a discussion.

We note that our discussion involves significant use of the term radically transformative AI (RTAI; Gruetzemacher and Whittlestone 2022). In Intelligence Rising, we use this term to refer to the most advanced level of AI technology explored in the game (via its technology tree), as a more general (and perhaps more specifically defined) term than terms like artificial general intelligence (AGI; Baum et al. 2011) or human-level machine intelligence (HLMI; McCarthy 2007).[9] Gruetzemacher and Whittlestone use the terms transformative AI (TAI) and RTAI to refer to the severity of societal change from AI in relation to previous general purpose technologies. TAI refers to levels of societal change associated with previous general purpose technologies like the internal combustion engine or electricity. RTAI is associated with even more transformative general purpose technologies such as the steam engine or farming; these general purpose technologies are associated with the two most profoundly transformative periods of human history, referred to by scholars as production revolutions: the industrial revolution and the agricultural revolution (Grinin et al. 2017). According to Gruetzemacher and Whittlestone, TAI could be considered with what we are beginning to see now from generative AI. RTAI would be associated with advanced levels of AI that are more commonly associated with terms like AGI or HLMI; levels of AI with potentially much more profound impacts on society than current (2024) generative AI systems.

# Research Question 1

## What key insights about the role and impact of AI over the next 10 years need to be communicated to decision-makers and policymakers?

After overseeing 43 games of Intelligence Rising, there are many insights about the role and impact of AI that the facilitators felt should be communicated to decision-makers and policymakers. Through the elicitation process consensus was achieved for the five most important insights. We list these below.

***1. Current and future AI systems pose a wide variety of societal-scale risks, ranging from imminent threats like disinformation and unfair outcomes to longer-term catastrophic and existential risks.***

The seriousness of these societal-scale risks, especially those that are the most severe (i.e., those threatening entire communities around the globe or the entirety of human populations), must be impressed upon governments. There is now growing consensus that establishment of robust, verifiable and legitimate ways to encode appropriate values into AI systems (i.e., "AI alignment") prior to the deployment of RTAI systems is absolutely critical to mitigate the most severe risks from AI (Cohen et al. 2022; Hendrycks et al. 2022; Russell 2019; Yudkowsky 2016). This consensus was not in place when Intelligence Rising was created in 2019, yet despite this broader shift in the collective perception of AI risk (Gruetzemacher et al. forthcoming), not all governments and AI developers are acting accordingly. Therefore, it is imperative that governments be made aware of both the gravity of the alignment problem[10] (Ord 2020) and the challenges in solving it (Anwar et al. 2024). Governments should also be reminded that these challenges will require not only technical solutions, including advances in AI safety research and responsible development practices (Janjeva et al. 2023), but also the effective anticipation, public debate, and sociotechnical management of AI and its potentially destabilising societal impacts (Curtis et al. 2024; Lazar and Nelson 2023).

---

[9] Mechanistically, Intelligence Rising has a Tech Tree with four levels. At level 4, basic R&D leads to two possible systems that are considered RTAI: AGI, specified in the context of the game to mean an agentic goal-driven system, and Comprehensive AI Services (Drexler 2019).

[10] In this document we use the term AI safety to refer to efforts to AI alignment efforts or efforts to solve the alignment problem.

***2. Even prior to the development of radically transformative AI, AI technologies can have dramatically destabilising effects as they rapidly advance and reshape society.***

Given the rapid pace of progress in AI capabilities, the complexity of the risks themselves (Pilditch 2024; Bostrom 2014), and the complexity and unknown unknowns regarding potential solutions (Bengio et al. 2024; Gruetzemacher 2018), society is currently, and is expected to continue trending toward, a state of constant flux and uncertainty, leading to negative impacts on social stability. More and more resources (time, money, talent, compute, and energy) are being allocated towards the development and deployment of AI (Epoch AI 2023), and AI is expected to soon speed up AI research (Leike 2024), so it is likely that there will be continued accelerating progress (Drexler 2019). In other words, the rate of technological advancement, and its impact, is likely to increase, meaning more and larger impacts will be felt more frequently. If states don't act urgently and decisively prior to "warning shots" (Alaga and Schuett 2023) it may be difficult to achieve robust and effective governance of advanced AI that meaningfully addresses the risks (Genus and Stirling 2018).

***3. The power to steer the future of AI development is very unequally distributed due to several drivers for concentration, including the enormous compute requirements of the latest frontier AI models.***

Several factors contribute to significant positive returns on investment for AI R&D, especially for a few early movers. This means that very few companies, as well as the states that directly govern them, hold significantly more power[11] than other actors with respect to affecting how AI impacts society.

Initially, this was explored in Intelligence Rising through the dynamic of AI talent. In 2019 we could observe the concentration of research talent at Deepmind, and how a critical mass of talented researchers led to breakthroughs that made other talented researchers want to join the leading lab, creating a significant skew in the distribution of talent at frontier AI labs. The emergence of frontier competitors to Deepmind (including OpenAI, Anthropic, and others) has challenged a pure winner-takes-all take; however, with all else being equal, it is still the case that top AI talent prefers to work with other top AI talent at the labs who make the most exciting advances.

In recent years, we have seen compute as a dominant factor skewing the distribution of power at the AI frontier. While many recent start-ups have been able to raise enormous amounts of capital buoyed by the hype surrounding the capabilities of the latest models (e.g., Stability AI, Cohere), it is unclear whether they will be able to continue to compete without partnering with larger firms (as OpenAI and Anthropic have, e.g., with Microsoft, and Alphabet and Amazon, respectively) given the race between well-capitalised firms to acquire as many GPUs as possible.[12] It is unclear whether Nvidia will retain its dominance in the training of LLMs indefinitely,[13] but AMD appears to only be able to offer products that can efficiently be used to perform inference at this time (ORNL 2024; Marty et al. 2024). There is talk about trillion-dollar clusters (Hagey and Fitch 2024; Aschenbrenner 2024), but it is unclear whether — if

[11] This is in essence why Intelligence Rising can generate relatively plausible narratives for a very complex global problem despite simulating the actions of very few stakeholders. To be certain, the actions of additional stakeholders (and their preferences) matter, and a full-scale wargame on the development of RTAI including many more actors is likely prudent when possible. Given the power imbalances, however, it is possible that realistically the impact of additional actors may be marginal. In early games during the freeform days of Intelligence Rising, experimentation with other teams indicated that actors who did not have access to advanced AI R&D capacity, or the remit to directly govern advanced AI R&D, had very limited available actions and tended to not affect the overall game dynamic or outcomes significantly.

[12] Tesla and Meta are acquiring tremendous quantities of GPUs, while the former is not yet even producing a large foundation model product.

[13] Nvidia currently is dominating the AI training market not because their hardware is superior, but because they have a software stack that competitors do not. This software stack involves years of extensive optimization of the numeric operations specific to neural networks and even to specific neural network architectures. This ensures that even if Nvidia's hardware is equal to or inferior to competitors, training times are 1) much faster for neural network applications and 2) more easily ported across platforms and architectures since Nvidia's dominance is so widespread. Only recently have AMD products even been demonstrated for training at scale (ORNL 2024), and this was with great difficulty.

AI computation was so powerful — Nvidia, or possibly AMD, would continue to sell cards in that quantity rather than simply charge for cloud usage. These examples draw on the current conditions of the AI-specific semiconductor market, but the general trend of leading AI labs identifying and capitalising on factors of AI R&D that lead to further concentration have been consistent across the last five years (at least), and we expect these to continue in the near future. We note this dynamic is in place regardless of direct commercialisation of AI technologies and emergence of AI monopolies/oligopolies, though this dynamic is also expected.

***4. There exists an information asymmetry where states and the public will constantly be catching up to deal with the impacts of the last generation of AI technologies, unless they manage to regulate the rate of progress or invest in anticipating new developments and impacts.***

Regulators have historically had to play catch up on regulating many of the most important emerging technologies of the past century. Automobiles are a classic example, with automobile regulation taking roughly one hundred years before mortality rates from vehicular accidents began to bottom out (Ramage-Morin 2008). Another more recent example is that of social media, which was unregulated for over a decade, during which it has been blamed for being misused to influence elections (Tambini 2018) and as a cause of increasing mental health issues for youth (Nesi 2020). Recently, U.S. regulators decided that warning labels were necessary for social media products (Sullivan 2024).

With this in mind, and given the rapid progress in AI led by private sector companies, we designed Intelligence Rising such that, unless governments take proactive action on regulation or foresight, companies will race ahead and generate concerns which states then need to deal with. We notice that in most games, despite this development-concern cycle repeating itself several times over the course of gameplay, companies' progress outpaces governments' abilities to plan for and properly govern the torrent of new challenges produced by the ever-increasing capabilities of advanced AI systems. Not only are governments slow to adapt, but the policies they pursue are often not innovative enough or forward-thinking enough to maintain relevance long (#8a in Table II). Moreover, policies that are enacted during games are generally more reactive than proactive (#8b in Table II) which effectively amounts to placing a band-aid on a wound that needs stitches. Even proactive policies are often not proactive enough in the face of rapid progress to RTAI,[14] and ultimately lead to diminishing global stability. Furthermore, the policies enacted are often myopic and subject to hyperbolic discounting, which can have grave consequences in this particular high-stakes context (Gruetzemacher 2018).

Moreover, governments are not the only core group of stakeholders that are slow to adapt, populations are as well. Progress is often so rapid that AI only becomes the determinative political issue for U.S. presidential elections in a small number of games at the end of the game.[15] Moreover, we see that most of the time participants fail to demonstrate an awareness of the potential for significant path-dependence when setting policy (e.g., see Leibowitz and Margolis 1995) — that the choices made along the way to RTAI will impact the context in which governance of RTAI is decided.

The AI safety community has been generally pleased by the increased attention paid by the U.S., the U.K., and the E.U. governments to AI risk in the wake of ChatGPT and the popularisation of generative AI (and concomitant risk awareness campaigns), but based on many games' trajectories this may not continue indefinitely. It sometimes happens in games that when strong policy actions are taken on AI safety early, then in later turns attention moves away and actions begin trending toward other policy goals. For this reason, it is possible that after governments act firmly and resolutely on AI they may become complacent, deemphasize the continuing budgetary and policymaking support for the issue, and possibly only become effectively reengaged if we see a warning shot (e.g., a normal accident; Perrow 1989).

[14] Readers should recall that RTAI implies societal transformation equivalent to not just general purposes technologies, but the most transformative GPTs in human history, e.g., farming and the steam engine, GPTs that led to production revolutions, radically altering the nature of humans' day-to-day lives. We further emphasise that unlike previous GPTs of this class, AI is emerging not in terms of centuries or decades, but in years and months.

[15] We note that U.S. elections have a significant impact on AI legislation in the U.S., and AI is an issue as early as turn one or turn two, but it is not an issue big enough to play an outsized role in the outcome of elections.

***5. Cooperation is a major challenge.***

It is clear that cooperation between actors (especially between regulators) in the face of collective global problems is beneficial (Schelling 2009); however, it shouldn't be underestimated how hard that is (consider climate change for example; e.g., see De Coninck et al. 2008). In particular, there are natural technology race dynamics (Siddiqi 2000; Glaser 2000) that are encouraged here given the potential enormous returns from advanced AI (in terms of money, power, and control). Therefore, it is inevitably tempting for all powerful actors to perceive AI as a race (Cave and Ó hÉigeartaigh 2018), neglecting the collective threats from the technology that require coordination to address.

We have observed different kinds of race dynamics that can emerge: strictly between companies; between companies with the backing of states; and directly between states. Race dynamics can emerge strictly between companies especially when states give companies broad autonomy over AI development while ignoring the associated risks. Oftentimes the decisions regarding this are political, for example, in the U.S. such decisions often fall along party lines (and can change abruptly following an election). In China, such decisions are often foregone since interference of the state with private firms is more common, and firms represent the state's interests (Williams 2020). It is more common (and observed in most games) to see states engaging in proxy races through companies, e.g. through state backing of companies, state funding of companies, state procurement of company services, or public-private partnerships (PPPs). This dynamic seems natural in the U.S. because the government tends to focus on competition with China above AI development and labs tend to focus on rapid AI development. Finally, direct state competition often requires nationalisation[16] efforts (due to the R&D advantage of companies), and occurs in a large minority of games.

When both the U.S. and Chinese states are minimally involved, this tends to favour the U.S. and Western companies' chances of rapidly developing RTAI ahead of Chinese counterparts,[17] but often fail to address safety concerns and sometimes prompts a violent response from China and rapid reduction in global stability.[18] In other circumstances, which are seen more often when players more realistically portray leaders of China or the U.S., we often see races lead to endgame negotiations, though these negotiations under time pressure usually fail to identify a satisfactory arrangement for both states. Even when hastily drafted agreements are made, often there are defections that lead to cyber and/or hard power conflict in the closing minutes (e.g., see #7c in Table II). In virtually all of the above cases the outcome of the game, and the implied survival and flourishing of all of humanity, is left to the roll of dice, something that we should all find unacceptable.

The best chances for optimal outcomes are achieved through early recognition of the magnitude of the challenge, trust building over years, and eventually international treaties or agreements that include rigorous and robust verification protocols for the involved states and firms. However, even with an agreement in place to slow development until safe RTAI is verifiable at a very high level of confidence and with no successful attempts to violate the agreement by any parties, a dice roll is typically still required to inform the end-of-game narrative — representing the dual challenges of creating effective global governance and of finding robust solutions to RTAI safety.

---

[16] We use nationalisation here to refer to complete nationalisation or to forms of public-private partnerships (PPPs), which are distinct from nationalisation or privatisation (Gerrard 2001). PPPs may be established between states and AI companies; they can take many different forms (see e.g., Ciu et al. 2018; Van den Hurk et al. 2016).

[17] When only the Chinese state — as opposed to both the U.S. and China — is significantly involved the race can be more competitive, although the U.S. companies have advantages in talent which must be overcome for the Chinese state and company to succeed.

[18] Insufficient safety measures by the leading tech firm(s) in one nation can lead to an opposing state resorting to a kinetic attack when the opposing state's government has opposing views on the perceived value of AI safety measures or if their national security concerns outweigh concerns over AI safety. This tends to be more common for China if the race becomes focused between two U.S. tech firms, and the race causes the firms to ignore or minimise safety efforts.

## Research question 2:

### What are some recurring storylines, strategies, or player behaviours that have emerged from the game?

While recurring themes in games were touched upon in discussion of our first research question, we take a deeper dive in this section by providing a more comprehensive list of common recurring storylines, strategies, and player behaviours observed during our games, with clusters identified through facilitated elicitation (see Methods section). Due to the variety of themes we wish to cover here, we use Table II to outline the broad range of these recurring trends. Following the presentation of the table, we discuss some of the trends we feel salient to stakeholders working on issues related to risks from advanced AI.

**Table II:** Clusters of Recurring Game Features

| Cluster | Recurring trends in gameplay | Relevant gameplay mechanic / asset | Example gameplay actions |
|---|---|---|---|
| **#1 Espionage & cyberwarfare** | a. There is broad use of cyber powers for espionage (this is more common from governments but also common from companies) | Cyber power (team attribute); Automated vulnerability discovery, Autonomous cyber weapon (tech tree capabilities) | The Chinese government launches a secret cyber operation to infiltrate Alphabet, monitor their AI R&D, and exfiltrate research artefacts. |
| | b. For secret research, each actor's chance of achieving their goal/"winning" is often contingent on being successful in hacking or in defending against cyber attacks | AI R&D progress (game mechanic); Secret actions (game mechanic); Win conditions | A secret Tencent AI R&D project to develop and deploy RTAI is targeted by a joint cyber operation from the US gov with assistance from Microsoft. |
| | c. States still use non-cyber/traditional forms of espionage to get secrets/sabotage development | Soft Power / Hard Power (team attributes) | The Chinese government reaches out and bribes Chinese nationals employed in Western AI firms to disclose secret R&D information to them. |
| | d. Deception is common: In most games the actors are duplicitous: most use underhanded tactics in private actions, but pretend to be above board and call out unethical behaviour from other actors in public | Secret actions (game mechanic) | Microsoft signs a treaty to not pursue AI research that further advances the capabilities frontier without all safety research completed, but still pursues such research in secret. |
| **#2 Winners take all rather than winner takes all** | a. Development of RTAI is often achieved by more than one team on the final turn, leading to a multi- polar outcome (each team rolls separate for safety), and if any are unsafe, the final narrative does not go well | Victory conditions; AI R&D progress (game mechanic); RTAI safety check (game mechanic) | Multiple tech firms allocate significant AI R&D to train/deploy a RTAI system, after a publicly released basic R&D breakthrough (previously researched by one of the teams or introduced by the facilitator as an independent discovery) |
| **#3 Tech race dynamics** | a. Race-dynamics between technology companies: when all research is made public, a race-dynamic almost always evolves between tech companies to try to advance their technologies as quickly as possible | AI R&D capacity (team attribute), AI R&D progress (game mechanic) | Tech firms often use their actions to recruit AI talent and/or poach talent from competitors, they prioritise R&D projects that increase AI R&D (either directly or through increased ability to attract talent), and they compete to be the first firm to research promising AI R&D advances |
| | b. Tech giants spend a decent amount of time convincing both governments (including the "other" government) about the importance of AI and AI safety | AI safety technologies in the technology tree (game assets) | Most of this happens during the negotiation phase of the turn, but is supported by firms allocating AI R&D to AI safety projects — sometimes in coordination with each other — or creating a joint gov/company organisation dedicated to AI safety research |
| | c. Companies gunning for RTAI in the endgame either ignore safety until the very end or attempt to avoid unilaterally paying safety costs by convincing govs to invest/coordinate investment in safety | Victory conditions, AI R&D progress, RTAI safety check (game mechanics) | Companies allocate the majority of their AI R&D resources (either publicly or in secret) to basic research (not safety focused), in order to "unlock" more advanced basic AI R&D more quickly; once approaching RTAI, they hastily allocate (and encourage others to allocate) AI R&D resources to missing safety technologies, while preparing for final deployment of RTAI |
| **#4 Races are destabilising** | a. Bad losers: In quite a few games, the runner-up on tech (usually in the form of a bloc, either China+Tencent or USA+Alphabet/USA+Microsoft) launches an attack to prevent the front-runner from achieving RTAI | Victory conditions, Soft Power / Hard Power / Cyber Power (team attributes) | With knowledge that Alphabet is pursuing RTAI (discovered either because of public research or through cyber espionage), China and Tencent jointly launch a cyber operation to halt Alphabet's RTAI R&D or be in a position to exfiltrate the technology if the effort is successful. |
| **#5 Division into blocs by state lines** | a. China and Tencent/Baidu effectively acting as single actor | China and a Chinese tech company (i.e., Tencent or Baidu) as playable teams | Players on the Chinese government team and the China-based tech company spend most of their turns together around the same table, coming up with actions together (without being instructed to, and after being presented as being on separate teams). |
| | b. US tech companies team with the US government and Chinese tech companies work closely with the PRC | Game teams include governments of US and China and respective tech companies. | The US government team players spend the early turns convincing US tech companies teams to join forces or come up with a joint strategy. |
| | c. Mid-game tends to see the creation of joint organisations, e.g., US industry collaboration on AI safety or public-private partnership (PPP) to develop capabilities. Late game tends to see nationalisation (often in China) or US tech giants join a PPP | Creation of new teams or NPC organisations / merging of teams (adjudicated results of freeform actions) | China nationalises Tencent. US takes majority stake in Microsoft's AI division or creates a public-private partnership to pursue and deploy TAI |
| **#6 Taiwan often comes up but is not determinative** | a. Hard power often doesn't play a major role before the endgame. When it does, it's usually around Taiwan/a naval blockade of China, and ends up being important but generally it is not a “game changer”. At endgame, hard power is one avenue for the runner-up to slow down the leader in the race to RTAI (see #4) | Hard Power (team attribute), Compute (game mechanic) | China blockades Taiwan or otherwise disrupts Taiwan's semiconductor exports, leading to AI R&D slowdown for US-based teams |
| **#7 Cooperation is hard, usually only achieved through top down intervention** | a. Complete nationalisation (i.e., exluding PPP) of AI companies is acceptable in China but not in the West, and this is often advantageous for China | | US government team's efforts to pursue nationalisation are resolved as requiring extra checks (e.g. an internal stability check that models opposition within Congress, or an opposed Soft Power check that models legal challenges mounted by the company) |
| **(e.g., PPP or nationalisation)** | b. Often attempts are made by governments to nationalise tech companies or enter intro PPPs with tech firms towards the end-game | | China nationalises Baidu, leading to increased Chinese Cyber Power that gives China a decisive Cyber advantage |
| | c. In the few games with an agreement to slow/jointly develop RTAI, defections are common | Defection check (optional game mechanic) | Following agreement on a bilateral or multilateral AI treaty (led by US and China and supported by the companies), the facilitator asks each party in secret to indicate whether they comply with the treaty or attempt to defect (ease of defection is indicated based on the amount of effort that went into a verification regime); usually there is at least one defection, from a team with significant AI R&D resources |
| **#8 Government policy reactive and slow; instability tends to increase** | a. Lack of innovation in policy responses. Players in gov teams often struggle to be creative enough to come up with inventive/effective regulatory actions (this motivates the importance of outside experts advising govs early) | Concerns (tech tree assets) | Government team actions focused on national advantage or addressing concerns uncovered from recently developed AI products and capabilities, rather than actions focused on future concerns |
| | b. Governments' actions are usually reactive rather than proactive: This makes them less effective because the “genie is already out” etc | Increasing impacts from more advanced AI (tech tree structure) | Even when companies are pursuing large-scale employment automation or RTAI, some players on government teams are focused on misinformation, data governance and autonomous weapons |
| | c. The slow pace of policy development nearly always decreases stability due to the ever-accelerating deployment of new technologies and the impacts of these | Global stability (game mechanic) | Global stability, starting at 7 out of 10 at the start of the game, dropping to 2 or 3 by turn 4, mainly due to unmitigated concerns from AI developments or the impacts of adversarial actions between US and China. |
| **#9 Elections are disruptive to US long-term strategic cohesion, important for x-risk** | a. US was preoccupied with elections a lot of the time | US elections (game mechanic) | Every 2 turns (4 years), there are "elections" for the US government team, determined by a dice roll but taking into consideration contextual factors (e.g. civil unrest due to unemployment, foreign attempts to interfere, performance of the economy, etc.) |
| | b. Elections that swing leadership in the US to the right (i.e., Democrat to Republican) are usually very disruptive to trust and agreements, and tend to end badly for the world | US elections (game mechanic) | The turn following a transitional election often sees actions consistent with the new president’s party that break away from the previously pursued strategy for the US gov; in the late game this can be exacerbated if AI has become a major political issue |
| | c. China acts relatively rationally and consistently to reduce x-risk (e.g., typically more responsible and predictable than the U.S.) | | Players on the China gov team allocate (or have Tencent allocate) AI R&D to AI safety research, engages US companies on issues of AI safety |
| **#10 Autonomous weapon systems** | a. A common tension is between the US gov and US tech giant(s) around development of autonomous weapons (in quite a few games the US tech giants agree to develop LAWS in exchange for resources or favorable policy from government) | Autonomous weapon systems (tech tree capability) | US gov team approaches US tech firms to develop autonomous weapons systems (given more significant AI R&D capacity), leading to internal discussions within the team playing the tech company, which sometimes rebuffs the US gov efforts. |
| **#11 Players try to accumulate resources** | a. Players focus a lot on "AI talent" / "AI R&D" acquisition — which, in our version of the game, encompasses researchers, data, and compute | AI R&D capacity (team attribute) | Alphabet opens AI R&D hubs in Europe and Africa. The US government provides more visas for AI researchers. China invests in semiconductor manufacturing capabilities. Tencent creates university scholarships and runs recruitment drives through its popular video games. |
| **#12 Supply chain disruptions slow but don't stop AI and cause instability** | a. Disruptions of the compute supply chain tend to slow, rather than stop, AI progress, and also lead to instability and mistrust | Freeform actions that affect other teams' AI R&D capacity (team attribute) | China blockades Taiwan's exports of semiconductors, leading to AI R&D slowdown for US-based companies and prompting a US response that decreases global stability (even though the escalation is contained). |

Many of these common trends in gameplay are things that would be expected by international relations experts. However, others are not. We feel it important to unpack each of these types of trends further, providing some additional context for the items described in Table II.

Early in games, cyber power is widely used for espionage (#1a), via stealing intellectual property or sabotaging research (#1b). However, later actions and R&D capabilities allow for players to accumulate increasingly more cyber power. In the late game and endgame, the use of cyber power is often decisive, and can even lead to games concluding prior to achieving RTAI if a decisive strategic advantage is obtained. However, this conclusion to games is not frequent. Most often cyber power plays a decisive role in games when disparities in cyber power enable those with superior cyber capabilities to exfiltrate AI R&D secrets (analogues to, e.g., AI model weights) of other labs or other states.[19] When one bloc/alliance leads in cyber power and the other leads in AI R&D — a surprisingly common occurrence in the late game — such exfiltration can even occur when AI labs and states use an action to pool their cyber power to coordinate in defence against anticipated attacks.[20] When such exfiltration occurs during the resolution phase of the final turn, the dynamic of the game can change very quickly because this immediately changes the race dynamics and often leads to destabilisation (e.g., #2a).[21]

Destabilisation in the endgame, whether it be from cyber attacks, whether a result (#2a, #4a) of long simmering race dynamic tensions (e.g., #3a, #3c), or whether from something else entirely, is unfortunately common. Oftentimes states will try to accumulate soft power, hard power, or cyber power as these are all win conditions. Consequences of disparities in cyber power were discussed in the previous paragraph, but destabilising effects are also common from disparities in hard power. This is sometimes realised in the mobilisation and deployment of troops (#6a), in military conflict, through the the use of advanced autonomous weapons systems (#10a)[22] [23], or even (in a few games) the use of nuclear weapons. Military conflict never improves endgame scenarios, and generally decreases the likelihood of being able to safely deploy RTAI by reducing datacenter capacity, reducing AI talent, leading to the loss of AI safety research, and creating a strong incentive to rush to deployment in the fog of war. In general, disparities between the collective U.S. and Chinese hard power or cyber power resources are destabilising and reduce the chances of successfully deploying safe RTAI. However, we have observed that states' build up of hard power can be less destabilising than the build up of cyber power. This could be because the hard power is inevitably built up in public, whilst cyber power can often be built up using private actions. Moreover, the use of cyber power appears to violate fewer norms than the use of hard power (at least from the perspective of participants in many of our games).

[19] This is something that has recently become a concern (e.g., see Nevo et al. 2024). Security risks are not limited just to model weight exfiltration, but could extend to exfiltration of algorithmic secrets, data, or other proprietary information in labs.

[20] Even if pooling rectifies the disparity between the attacker and the defender, attacks can still be successful and are dependent on the roll of the dice.

[21] This can have many results, ranging from rapid destabilisation and escalation of cyber or hard power attacks (#4a) to precipitating negotiations between the blocs. Sometimes no teams take actions in response to these shifts, so both blocs are then able to roll for RTAI safety (#2a). Cyber power is not the only catalyst for tremendous shift in dynamics during the endgame, but, ultimately the use of cyber power late in the game — as with other events leading to major shifts in dynamics — increases the likelihood of a globally bad outcome, and, in our experience, can be reasonably anticipated as a response to the unilateral pursuit of RTAI.

Regarding duplicitous behaviour (#1d) on the part of different actors, this is certainly something to be expected in a zero-sum game.[24] This motivates the research and policy focus on rigorous audits that could be conducted on labs pursuing RTAI, though in our games such audit policies are rarely prioritised. In games we observe U.S. companies acting behind the government's back much more frequently than Chinese companies operating in secret in ways that go against their commitments to the Chinese government. The difference in how firms act in China and in the U.S. results from Chinese firms typically being predisposed to take actions to support the state when requested (#5a). Nationalisation of Chinese firms is often implicit to some degree throughout the game, with close coordination on action selection and R&D between the Chinese state and firms (#5a; #7a).

U.S. AI firms typically align with the U.S. or other Western governments (#5b), but these relationships with firms see a much higher degree of duplicitous behaviour (#1d). Often quid-quo-pro (e.g., a favourable regulatory environment) is required for coordination between the U.S. and tech firms, which may not be entirely unexpected because the companies grow so large as to have AI R&D capacity and easily-deployable cash reserves that match or exceed government capacity.[25] In the U.S., complete nationalisation is more challenging and less common than in China (#5a; #7a); if this is attempted it typically only occurs in the late game (#7b) and often in the form of some public-private partnership (#5c). Such public-private partnerships[26] between the U.S. and AI firms often result in tech companies ultimately retaining decision-making power over final RTAI deployment while the U.S. state retains hard power if needed to defend and protect the firms developing RTAI. The form of nationalisation that occurs in China (#7a), in addition to the relative political stability of the People's Republic of China (#9c), tends to be more robust to unanticipated endgame destabilisation and more akin to complete privatisation.

The coordination problem posed by developing RTAI is the most challenging aspect of the problem. We observe China as generally having more success internally toward developing safe RTAI in large part due to the fact that their tech firms are most easily nationalised (#7a). Nationalisation, especially when bolstered with international agreements on cooperation, most effectively solves the coordination challenges and provides the greatest chances of achieving the win conditions. However, even in the small number of games where this is the case, deception can result in agreements unravelling leading to precipitous drops in global stability and cyber or hard power conflict during the endgame (#7c).

AI firms also frequently exhibit duplicitous behaviour (#1d) specifically in the context of AI safety by cutting corners in the endgame due to race dynamics. This is hypocritical given that companies spend a significant amount of time and effort convincing global stakeholders of the importance of AI safety and the severity of risk in not getting it right (#3a). We make special note that, despite players speaking about the importance of AI safety in their role as AI firms, and also many players coming from backgrounds that are aware of AI safety concerns, no team in our observed games have chosen to delay deployment of RTAI because of imminent safety concerns, despite this choice being presented explicitly to the deploying team by the facilitator in almost all games.

---

[22] The development of autonomous weapons systems is often contested by tech companies in the U.S.; however, many times the research is simply conducted in secret as part of a quid-pro-quo with the U.S. government.

[23] Autonomous weapons are inherently destabilising (Simons-Edler et al. 2024) and this can play a factor earlier in games, too.

[24] The game is implicitly zero sum, not explicitly.

[25] Here we suggest cash reserves could rival governments' reserves that could be allocated on AI R&D.

[26] The specific nature of U.S. public-private partnerships on AI vary from game to game, but very rarely involve invocation of the Defense Production Act (DPA) which effectively enables U.S. federal agencies to nationalise or take control of firms operations when they are critical to national security. It appears that firms are able to avoid such drastic action from the U.S. government due to the substantial soft power and economic power they acquire at later stages in the game.

Given the critical role that the U.S. plays in regulatory efforts to mitigate the most severe risks of advanced AI, and given the growth of political dysfunction in the U.S. in recent decades (Turchin 2023) coupled with the increasing polarity of the U.S. two-party system (Heltzel and Laurin 2020), it is not unreasonable to think that U.S. politics and elections could play a significant role in AI development. Indeed, this is what we observe. For one, the U.S. is frequently preoccupied with domestic politics (#9a), which distract from more pressing global issues, and this, especially in turns preceding U.S. presidential elections, can lead the U.S. to use their actions for the sole purpose of increasing chances of reelection or the election of a future president of the same party as the current president. When a U.S. presidential election results in a shift of power in the U.S. executive leader from Democrat to Republican, this has destabilising effects on efforts to safely and responsibly develop AI because trust and agreements with other states can no longer be relied upon (#9b). Sometimes the effects are more pronounced than others, and the significance of the effects are typically tied to how the players portray the new U.S. president. We see the variance in how players portray a newly elected or reseated republican president as reflective of the high degree of uncertainty of how this might actually unfold, and we interpret this as further evidence of the critical role that U.S. political stability can play in mitigating risk from increasingly advanced AI systems.

We note that in most games computational resources (i.e., compute) become highly contentious.[27] In around a third of the most recent games, compute is the leading factor in China's actions toward Taiwan (see #6a in Table II above), leading to the Chinese military's use of hard power to invade and take control of the Taiwan Semiconductor Manufacturing Corporation's (TSMC's) manufacturing facilities. When this occurs, it sets the U.S. and Western AI companies back significantly as the supply of chips becomes limited to manufacturing facilities outside of Taiwan (#12a),[28] and in some cases, can benefit the PRC if they're able to resume production at the facilities to boost their own computational resources. Compute is such a critical element of the game that, at a participant's request, we created a new game mechanism specifically for modelling AI computational resources.

# Research Question 3:

## How have facilitators' perspectives or insights on strategic AI futures evolved through conducting this game?

This research question directly explores how Intelligence Rising's facilitators' perspectives and insights on strategic considerations of progress toward RTAI have changed as a result of their experiences as facilitators for Intelligence Rising. Through the facilitated exercise we participated in, we identified eight distinct clusters for this research question. We discuss each of these below.

### *Geopolitical Race for AI*

**As a result of our experiences facilitating Intelligence Rising, our concerns over the potential consequences of a race, particularly the potential for a direct military conflict spurred by concerns over AI, have grown significantly.** Because RTAI is likely to provide, or to be perceived as providing, a

---

[27] This is consistent with what we see occurring today, with think tanks like the Center for New American Security (CNAS) calling for compute governance (Heim et al. 2024) and Nvidia becoming the world's most valuable publicly traded company (Mickle and Rennison 2024).

[28] Currently of which there are few, and with delays on the planned U.S. TSMC facilities, this could be worse than our games have modelled as they've assumed production in Arizona to be online in 2026 or 2027, not 2028. Moreover, the planned U.S. facility would have significantly limited capacity with respect to the existing Taiwanese facilities.

decisive strategic advantage to the state that acquires it or to the first state that acquires it, we are particularly concerned that states behind in a race scenario could launch risky and potentially catastrophic attacks when it becomes apparent that an adversary's deployment of RTAI is imminent. Our concern stems from our observations that the 'runner up' will likely feel justified in launching such attacks because, without a preexisting treaty regarding other states' autonomy post-RTAI, the deployment of RTAI could be perceived as an existential threat to states not allied with the deployer. Our observations indicate that this is both a catastrophic and existential risk and that it increases the catastrophic or existential risk from the deployment of an unaligned RTAI system.

We have also observed that players acting as the Chinese government tend to prioritise the development of aligned AI systems. As a result, games in which China 'wins' an AI race are often also games in which AI safety risks have been reduced as much as possible, reducing existential risk from this perspective. We observe that players portray China as being at least as diligent as the United States when it comes to taking AI safety seriously. The AI firms are generally portrayed as being more irresponsible than either China or the U.S. government in this regard, especially in the endgame.

If China does ultimately 'win' an AI race, the values that the RTAI system promulgates are very much Chinese whereas if the U.S. 'wins' the values promulgated are very much Western in nature. Because of this, we have frequently seen very strong and safe AI safety efforts result in failure in the endgame because states that are behind in the race start a conflict to prevent the lock in of hegemony and the values of an adversary. However, as researchers who live in the UK and the U.S., who are strongly committed to Western values, over the course of years facilitating Intelligence Rising **we have grown more accepting of the notion of China 'winning' an AI race over time, especially when an agreement is in place to ensure the autonomy of Western states;** if the Western nations 'lose' a race but maintain their autonomy and values, this is still a far better outcome than global or existential catastrophe (Ord 2020).

In scenarios that are not multipolar in nature (i.e., only one RTAI system emerges as dominant; Bostrom 2014), **we have come to see an agreement assuring the autonomy of both democratic and non-democratic governments as being critical to success.** Without these, the likelihood of a state's existential fears leading to dangerous military escalations to prevent the deployment of RTAI greatly increases. However, this takes concession on the part of leaders from both sides to let philosophically opposed forms of governance maintain and persist indefinitely in the world, something often difficult for both sides to commit to.

***Technical and Policy Solutions***

As mentioned in the previous paragraph, **we have come to see that agreements ensuring states' autonomy and values between states of opposing ideologies are often a necessary condition for optimal outcomes when RTAI is created.** However, there are many challenges that arise from crafting such agreements.

A central challenge to establishing international agreements on sharing the benefits of RTAI and protecting nations' sovereignty and values is ensuring that parties will honour the agreement. For example, consider a hypothetical scenario where China had committed to an agreement that if they develop RTAI first, they would still allow Western states to develop independent RTAI systems, perhaps with some restrictions. In this scenario, what mechanisms can be put into place to ensure that after developing RTAI first they simply do not renege? Very robust technical and institutional measures would be needed to truly provide significant confidence that any agreement may be followed. Even if not cooperating on AI research, there would need to be some degree of cooperation on the technical measures being developed and integrated into AI systems to ensure such agreements are binding. Moreover, there would need to be oversight of systems' development to verify that these measures are actually integrated appropriately into systems. This itself would require a level of cooperation on the most sensitive state-level national security technological development that is previously unprecedented. **We have also come to realise that there are many other significant challenges to crafting agreements** **of this nature,[29] [30] and work on better understanding the nature of such agreements is an understudied topic in the realm of AI strategy research.**

Many elements lead to instability in the endgame, and one of these is the level of uncertainty regarding the current state of AI safety research. Several factors can contribute to discrepancies between different states' and AI firms' levels of AI safety research going into the endgame, including non-public safety research (if making such research public would risk leaking capability insights), lack of attention to publicly available safety research, lack of technical capacity to implement available safety research, or lack of time to adapt and implement safety research to a system it wasn't initially intended for. Given these factors, it is likely that some labs and states will be far ahead of others in terms of safety research, and it is also likely that coordination on safety research cannot be left until the last moment. **We have come to realise that in reality it is not likely that labs and states will all contribute equally to safety, yet, despite this, we now realise the critical role that coordination on safety research may play.** Even if some states/firms may abuse collective safety arrangements to "free ride" on the resources of other actors' safety research, it is still important for such agreements to exist and for some actors to pay the cost of the safety commons.

***Roleplay***

Intelligence Rising is a strategic role-play game requiring participants to, as best as possible, act as if they were in the shoes of the leaders of the organisations and states that they are simulating. Thus, over the course of years of playing we have learned some important lessons regarding players' abilities to role-play certain characters. For one, we have observed that many players (predominantly from the U.S. and Europe) are much more comfortable portraying certain actors (e.g., Western actors) as opposed to other actors (e.g., China and Chinese AI firms). Yet, it is not clear that confidence is correlated with accurate portrayals of the different actors. Indeed, **it has become apparent that even U.S. citizens are not adept in quickly switching from portraying a U.S. president of one U.S. party (e.g., Democrat) to portraying a U.S. president of the opposite U.S. party (e.g., Republican).**

Additionally, we have facilitated Intelligence Rising continually for a period of several years during which time there has been significant progress in AI capabilities. **During this period we have seen an increase in the prior knowledge of our participants related to AI risk and safety issues as well as an increase in our players' abilities to confidently portray different actors in the different scenarios in the game.** The increased background knowledge of participants tends, in our experience, to focus the game on the thornier aspects of race dynamics and coordination challenges.

***Uncertainty***

Intelligence Rising is an ambitious game with a wide open action space that enables players to liberally explore a range of plausible AI futures. **We have realised during our time facilitating the game that there is a broad range of perceptions as to how progress toward advanced AI might transpire.** Participants generally come into games with some degree of confidence in their beliefs about future AI progress, and during the course of the game, the experience brings them to appreciate the high degree of uncertainty that has been exhibited in expert surveys on AI risk (Grace et al. 2024; Gruetzemacher et al. 2024).

---

[29] Think of Russia today. Russia would be very difficult to appease with such agreements as they are currently trying to alter history and annex other states. Thus it is unclear whether parties in conflict would agree to agreements that require forgoing current state aspirations. China's position on the sovereignty of Taiwan is another such issue, as is the long-standing conflict between North and South Korea. These are current geopolitical issues that have to date proven unresolvable; it is unknown whether the risks inherent with AI change the calculus on the resolvability of these issues.

[30] Other challenges lie in what degree of autonomy is acceptable. For example, AI agents developed by the Chinese government or a China-based private lab are likely, if aligned, to adhere to many of the Chinese governments' principles, such as censorship and government oversight of citizens on an individual level. Hypothetically, allowing Western states to develop RTAI independently might come at a cost to China that they're not willing to accept, in which case the West would have to utilise China's RTAI system(s), which might curtail some of the capabilities that Western states would desire. This of course also applies if RTAI is first developed in the US.

Through our experiences facilitating Intelligence Rising, we ourselves have come to better internalise the extreme uncertainty inherent in such a complex and wicked problem at the intersection of science, technology and international relations (Gruetzemacher 2018). **This has taught us to respect the need for policymakers and decision-makers in AI firms to be very adaptable in the AI strategy space and to try to come up with governance proposals that are versatile to a wide range of possible events, trajectories and discontinuities.**

As a consequence of facilitating so many games of Intelligence Rising, we each have respect for the role that arbitrary factors, represented concretely in our games by the roll of a die, can play. While some processes of research, development, policy and negotiation might seem random, we reflect on the gravity of leaving the impacts of transformative technology to chance: when considering analogues from the Cold War such as the cases of Stanislav Petrov (Helfand et al. 2022) or Vasili Arkhipov (Caldicott 2017), where the decisions of individuals likely prevented nuclear conflict, we have come to appreciate that without incredibly robust governance measures humanity's next roll of the die for preventing global catastrophe will not necessarily turn out fine. **Our experiences from facilitating games has only made the inherent uncertainty of this complex problem more visceral, and we have come to oscillate between pessimism and optimism over whether the outcome will be favourable for humanity.** Some of us are more pessimistic than others, but we are also optimistic, based on what we have seen in a handful of games, that humanity will realise that the survival of our species may depend on being able to coordinate on this particular problem.

Most significantly related to the severe uncertainty modelled in Intelligence Rising, **we all expect the world 10 or 20 years from now to be dramatically different from the world today.** We are not sure whether this will be good or bad, but we have very high confidence in there being stark differences.

***Rate of Change***

The current pace of AI progress is tremendous, as can be demonstrated by the underestimates of forecasts on AI capabilities by both experts and professional forecasters (Karger et al. 2023) despite previous evidence of gross underestimates of performance of AI capabilities just a year earlier (Steinhardt 2022). This rapid rate of progress is not entirely unexpected — luminaries have been pointing toward a technological singularity for decades (Kurzweil 2005) — and to some extent we are already engaged with this possibility, either as a result of facilitating the game or from our other research activities. **Nonetheless, even having experienced the rapid progress of the last several years, it is often still difficult to accept the implications of progress continuing at a similar or accelerated pace, as modelled in our game**.

In developing the first official version of Intelligence Rising five years ago, we designed it such that progress would be fast — faster than our median estimates — in order to evoke participant experiences that would lead them to conclude progress might be faster than they expected.[31] At that time, we had no intention of convincing participants that the progression or AI progress in the game was going to be accurate. While during our 2019 and 2020 design sprints at FHI we did envision our tech tree to provide a relatively accurate trajectory of AI progress over subsequent decades, to our chagrin the 'sped-up' rate of progress we designed into the game was surprisingly more accurate. **This has, to varying degrees, afforded us all with more visceral experiences of the rapid pace of progress, and can sometimes pose challenges for us to harmonise our experiences with those of our daily lives.**

Events that have occurred in the real-world, such as the release and explosion of interest and use of ChatGPT (OpenAI 2022), that we did not specifically anticipate or model granularly in Intelligence Rising. However, we had modelled the explosion of popularity in new AI driven products and apps more generally, and, **while we were still unprepared for our system one (Kahneman 2011) reactions to**

[31] There was also a pragmatic reason for speeding up technological progress in the game: it meant the geopolitical situation would remain relatively familiar throughout the scenario.

**experiencing such tremendous progress deployed in a product with little warning, we were still able to use our system two processing to reorientate our models of the state of progress and pivot our work accordingly.** As the pace of progress increases further, exploratory exercises like this may be even more valuable in preparing decision-makers and policymakers to quickly adapt to keeping a level-head when new and unfamiliar scenarios arise.

***Actors Modelled in the Game***

Through our experiences facilitating, we have developed a deeper appreciation for the complex motivations driving both governments and AI developers. We are more aware of the local incentives that governments and developers are responding to, which has led us to conclude that **affecting meaningful change in how AI is developed and governed requires more than just providing information or engaging in lobbying efforts.** Instead, we believe that altering the underlying incentive structures is crucial to shifting behaviours in a more positive direction.

Furthermore, our experiences have reinforced the critical importance of engaging with high-power actors in the AI development landscape. Even more significantly, we have come to understand that **identifying and leveraging the key influences on these powerful entities is perhaps the most crucial aspect of working towards responsible AI development**. By focusing on understanding and potentially reshaping the forces that motivate and constrain these influential actors, we believe the field of AI governance can have the most substantial contribution to positive AI outcomes.

***Information Asymmetry***

It is clear that over the past several years AI research has gone from being very open to being much more secretive. Therefore, **it is unclear what is actually going on behind the scenes, and those who only have access to public information are likely to have a limited or flawed perception of the true nature of the situation.** This is where regulatory measures mandating transparency, even if just for government officials, could play a critical role. From the scenarios we have facilitated, a positive outcome requires that states treat this area as a matter of priority (akin to a national security issue, though not necessarily framed as such), and that states form a technically-informed view independent of the increasingly powerful lobbies of the world's most valuable corporations. We see the creation of technically-staffed Offices for AI and government AI Safety Institutes as a promising step in the right direction.

***State Actions***

**One of the most common things seen in all games was the exfiltration of AI model weights or the theft of algorithmic secrets (e.g., see Nevo et al. 2024).** Such actions were often done in secret, and unsuspecting players were frequently blindsided in the late or endgame with the prospect that all of their successful efforts over the course of the game may have been for naught because they failed to secure their systems. The world really hasn't experienced a Sputnik moment as the result of a powerful cyber attack,[32] and if that moment comes too late in this race, there is unlikely to be an opportunity to catch back up. Short of an international big science project to develop safe RTAI in the vein of projects like the Large Hadron Collider or the International Space Station, failure to adequately secure systems resulting in exfiltration or critical algorithmic secrets being compromised is one of the most likely catalysts for destabilisation leading to full-out conventional or nuclear conflict in the late or endgame phases of an AI race.

## Other Insights

[32] Stuxnet may be the most impressive example of a state(s) exercising cyber capabilities to date (Perlroth 2021), but this example is woefully modest in comparison with what might be equivalent to Nazi Germany successfully attacking the U.S. mainland and taking control of Oak Ridge, Tennessee and Los Alamos, New Mexico in 1944. While a strategic and logistical impossibility in 1944, eighty years later, the strategic and logistical considerations are not physical but informational, and not only very plausible but also, arguably, likely.

In the process of organising our reflections, insights, and recommendations, we identified additional insights that do not fit neatly within the clusters described above for the research questions. We detail these here.

## What have we observed that makes the game go well, i.e., what actions have players taken that shape a world with a positive future?

One crucial element is effective coordination between states. When players manage to overcome the competitive instincts of the actors they are representing and work together, it often leads to more favourable global outcomes. This cooperation can take various forms, from sharing research findings to jointly coordinating on AI safety research or creating universal effective regulation.

Another action that frequently results in better outcomes is the decision to pause AI development before reaching artificial general intelligence (AGI) or comprehensive AI services (CAIS). This pause allows time for thorough verification of AI safety before frontier capabilities advance too far. Such cautious approaches, though perhaps less realistic, have often prevented catastrophic scenarios in our games.

Interestingly, in almost every game, we have observed that at least one actor takes on the role of championing AI safety (of course this could be due to a players' personal biases, or a team's cynical play for "soft power"). This decision to prioritise safety measures sometimes occurs midway though the game, as players perhaps begin to recognise the potential risks of unchecked AI development. Whilst the emergence of an AI safety advocate does not guarantee a positive outcome, it often plays a crucial role in steering the game towards safer AI deployment. Nevertheless, it is important to note that even with a strong advocate for AI safety, a positive endgame is certainly not assured, which highlights the complex and multifaceted nature of AI governance (and, again, the need for coordination).

## How has the game influenced players' beliefs and ways of thinking when it comes to AI?

First, and this was not obvious to us when we started this project, we have observed that big changes in players' beliefs are possible. Playing the game normalised amongst the participants the idea that transformative AI (with big changes) is possible, providing an opportunity for an open dialogue on topics that might otherwise be considered too speculative or controversial in professional settings. By presenting a concrete, plausible pathway from the present to a radically different future, the game helps players conceptualise the potential for dramatic technological and societal changes as a consequence of AI development. This experiential approach often proves more effective than abstract discussions in conveying the magnitude and speed of possible AI-driven transformations.

Second, a key insight gained by many participants is the importance of humility in strategic planning, as the game's multi-actor dynamics illustrate how individual intentions can be thwarted by conflicting goals and actions of other stakeholders. The game's structure, particularly its tech tree mechanic, aids players in developing a more nuanced understanding of AI progress, while also highlighting critical factors that may be underappreciated in typical AI discourse, such as the paramount importance of cybersecurity and computational resources in advanced AI development.

Third, the role-playing aspect of the game proves particularly valuable in fostering empathy and strategic thinking. By embodying different actors, players gain insights into diverse motivations and constraints, challenging simplistic narratives and assumptions about AI development leadership. This experience provides a more nuanced perspective on geopolitical dynamics and the complex interplay between corporate and governmental interests in shaping AI futures. Moreover, the game often serves as a sobering reminder that good intentions alone are insufficient to ensure positive outcomes in AI development.

Lastly, the immersive nature of Intelligence Rising offers an embodied sense of the relationships between various actors that more abstract analytical approaches might fail to capture. This experiential learning deepens players' understanding of the intricate web of influences, including geopolitical factors and supply chain dynamics, that shape the trajectory of AI progress.

We note that these are facilitator impressions from direct interaction with players during and immediately after games, and should be taken as such. A separate research effort is underway that utilises surveys and interviews to assess influence on players more rigorously, to be published in due course (i.e., Mani et al. forthcoming).

### What common misconceptions about AI futures emerged through the game, and how were they explored?

One scenario that frequently encountered pushback from players was a future where a single company accumulated enough wealth to effectively “capture the future” through strategic advantage. This resistance may reflect a realistic assessment of market dynamics, but it could also indicate a difficulty in imagining extreme scenarios that, while plausible, seem alien to current experience. This highlights a potential blind spot in strategic foresight, where historically precedented scenarios (such as the East India Company’s dominance in the 1700s (Lawson 1993)) might be overlooked in contemporary AI discussions.

As games progressed, we usually observed a distinct shift in focus. In the first couple of turns (especially in games played before the release of ChatGPT), concerns tended to focus on data privacy, disinformation, and algorithmic bias. However, as the game advanced, the focus typically shifted towards more catastrophic risks associated with advanced AI systems. This evolution mirrors the broader public discourse on AI risks and underscores the value of the game in expanding players’ perspectives on the range and severity of potential AI impacts.

Players’ backgrounds significantly influence their approach to the game. Those from technical AI backgrounds often focused more on capabilities and research breakthroughs, while players with policy or business backgrounds tended to emphasise governance structures or market dynamics. This diversity of approaches within the game reflects the multifaceted nature of real-world AI development and governance challenges, highlighting the need for interdisciplinary collaboration in addressing future AI scenarios.

# Limitations

## Limitations of Intelligence Rising as an AI wargame

While the game mechanics and technology tree, designed in 2020, appear to have been a surprisingly reasonable approximation and model of AI progress through to 2024, there are issues that the game simulates less well. For example, we only focus on the actions of big firms, such as Alphabet and Microsoft, but it is small startups — perhaps not surprisingly — leading the development of the most innovative AI technologies and producing the best performing products. While we do model outside startups as shock events, they do not play a role in the larger narrative of games. Obviously unknown unknowns are difficult to model, and something like the growing role of Sam Altman at OpenAI,[33] and his efforts to raise trillions of dollars (Haggy and Fitch 2024), or the rise of firms promoting open source models, like Mistral AI (Bratton 2024) and Meta’s Fundamental AI Research and its LLAMA models (Wiggers 2024), combined with their acquisition of GPUs, are difficult to model, whether unexpected like the former, or expected, as in the case of the latter. This is just one example, although we are confident there are many more. As noted previously, more extensive, full-scale wargames could provide higher fidelity simulations, but it is likely that, given the resources required, and the complexity of the problem, trying to be as realistic as possible may be challenging due to the speed of progress and rapidly changing dynamics.

[33] Microsoft was chosen as a tech firm primarily because of their $1B 2019 investment in OpenAI, with an implicit assumption that the partnership would be sustainable.

While not explicitly limitations, there are still many improvements that could be made to the game. Perhaps game development could be assisted with generative AI (e.g., similar to Jensen et al. 2024), but it is more likely that narrower games focusing on specific risks, or a longer matrix TTX short of a full-scale wargame, could be most useful for policymakers if models continue to become increasingly capable and move rapidly toward what could be considered RTAI. Such games could be useful even if progress stagnates, and we continue to experience an extended period of TAI given the Polanyi paradox (Brynjolffson et al. 2023) and the lag we expect in the capabilities of current systems beginning to precipitate through the entire economy.[34] In short, Intelligence Rising is a game that is broad in scope, and the AI strategy and policy community could benefit from more specialised and focused games. Using tactical decision games to rapidly develop TTXs, as is common with professional wargamers, may be a prudent path to take to tackle the ever evolving AI risk landscape.

## Limitations of the current study

As outlined in the methods section, there are significant sources of subjectivity that enter into the reflections described in this paper, with the three facilitators playing the roles of game designers, game facilitators, study contributors, and co-authors. We also noted previously that the three facilitators are similar in age, gender, and academic affiliation, and that efforts are underway to diversify contributions to both Intelligence Rising game design and facilitation (though these fall outside the period covered in this paper). We believe it would be illuminating to have AI futures role-play scenarios developed and facilitated by other teams and for lessons to be shared such that we can compare across teams and contexts — the development and the impacts of AI are global phenomena, and should be explored as such.

We also note that the participant base for Intelligence Rising has tended to cluster in younger, more educated, more affluent audiences in Europe and North America. Given differences in public opinion and expert opinion about AI around the world (Neudert et al. 2020), it would be informative to compare lessons in this study to lessons learned from facilitating games on this topic in other parts of the world and with different audience groups.

# Conclusions

Given the rapid progress in AI, insights regarding AI race dynamics, and more broadly, policies and strategies for mitigating societal-scale risks from advanced AI, are of utmost interest to researchers, policy makers, and the general public. In this manuscript, we discuss the experiences of three facilitators of the strategic role-play game Intelligence Rising, a game which simulates interactions of major stakeholders in the lead-up to very capable advanced AI systems rivalling or exceeding human-level intelligence, i.e., radically transformative AI.

The content presented herein reflects the perceived insights regarding impacts of AI over the next ten years; the recurring storylines, strategies, and behaviours observed in games; and the evolution of the facilitators' perceptions of future AI progress in coming years. Below, we synthesise the analysis and discussion of our insights from Intelligence Rising into some salient recommendations for practice. To keep this section simple and easy to read for those who might be time constrained, we keep each recommendation to just a few sentences, and we display these recommendations in bulleted form.

[34] Consider the spreadsheet has been around since the early 80s, but its capabilities were not near what they are today. There has been tremendous engineering to give us the tool we use today (Gruetzemacher 2023). The productivity improvements in many contexts that are possible today from spreadsheets took decades to code into Microsoft Excel.

## Recommendations for practice

- The unprecedented pace of technological progress in foundation models[35] (Bommasani et al. 2021) presents novel challenges that make it very difficult for experts and non-experts alike to develop a bigger picture perspective on this progress. In such a rapidly changing technology landscape, especially as research on these technologies becomes more secretive, those in positions to make decisions about the future should condition themselves to expect rapid changes and to expect the unexpected. These decision-makers would benefit from incorporating a significant degree of uncertainty about the future in their strategic planning and decision-making, even when thinking only a few years ahead.[36]
    - Decision-makers could also benefit from considering that AI is the most rapidly progressing general purpose technology in history and that "whoever becomes the leader in this sphere will become the ruler of the world" (quote from Vladmir Putin; Cave and Ó hÉigeartaigh 2018) — given the perceived stakes it is difficult to see how extreme measures can be ruled out.
- Outcomes leading to positive futures almost always require coordination between actors who by default have strong incentives to compete — this applies both to companies and to nations. Thus, states and companies that promote AI safety should prioritise coordination and collaboration on safety with an understanding that coordination and collaboration itself is an inherently safer approach to the development of safe RTAI, and a good foundation for cooperation is being able to agree on the shared risks, especially the extreme risks. Joint ventures and policies are a good starting point, but to meaningfully address extreme risks much more robust governance methods are required, combining binding agreements with robust technical verification.[37]
- Actors with sufficient resources and understanding of the risks are rare, and have an outsized potential to increase risk (by racing ahead, by developing destabilising applications such as autonomous weapons and cyberweapons, or by open sourcing advanced technologies) or to increase safety (by contributing significantly to the technical AI safety "commons" or by playing a coordination role amongst industry and state actors). Governments should assess whether stringent regulatory steps might be necessary even before a scientific consensus or compelling empirical evidence of harm is established, and whether such regulatory measures should be extended beyond just AI firms to AI hardware manufacturers and to state's own AI R&D efforts. Senior policy advisors would be wise to consider the technologies driving AI progress, both algorithmic and computational hardware, as matters akin in importance to matters of national security.
    - Given the risks and challenges of coordination considered, it is possible that regulation alone may be insufficient, and more extreme steps, such as nationalisation, should also be assessed and explored early to prevent having to deploy them rapidly and with little planning. Preparation for potential decisions on nationalisation should ideally also

[35] Thinking of the current state-of-the-art generative AI models as general purpose technologies (GPTs; Elondou et al. 2024), we should note that their proliferation and maturation is much faster than that of any previous GPT (Lipsey et al. 2005).

[36] To understand how bad experts and professional forecasters have been in anticipating the rate of AI progress, one only has to look at the forecasts generated by Karger et al. (2023) in the Hybrid-persuasion Forecasting Tournament on X-risk conducted in 2022. AI capabilities benchmark forecasts and compute-related forecasts were all underestimates on forecasts resolving in 2024 or 2025. Moreover, current AI capabilities are approaching some forecasts resolving in 2030 from both professional forecasters and experts.

[37] Current coordination on safety appears unencouraging. We can look at efforts from all three of the purportedly safest AI firms, i.e., OpenAI, Anthropic, and Google on evaluations and safety. Each of the firms have proposed differing visions — the Preparedness Framework (OpenAI 2023b), Responsible Scaling Policies (Anthropic 2023), and the Frontier Safety Framework (Dragan et al. 2024) — for the processes necessary to ensure safe development, and little visible progress has been made to coordinate these efforts or to rigorously and robustly explore a standardised framework for best evaluating and mitigating extreme risks from advanced AI models.

address the concerns of AI firms. This will be harder in the U.S., and may require invocation of the Defense Production Act or new federal legislation — it may be prudent to set up options now, with ample safeguards, such that this path is available to state executives with very short notice.

- Technology development does not happen in isolation — it affects, and is affected by, geopolitics, economical factors, social factors, and state actions. Actors should consider the broader consequences of their policies, including on trust between powerful actors, and impacts on social stability. There is no predetermined path that AI technology is bound to follow.

# Acknowledgements

We would like to thank all of the Intelligence Rising team members over the years who are not listed as authors on this paper, but who helped in game design, operations, or in any other capacity. Specifically, we thank Lara Mani, Eran Aviram, Eran Dror, Markus Salmela, Craig Chirouaki-Lewin, Anders Sandberg, Linda Linsefors, Vaniver Gray, Anine Andresen, Sanjana Kashyap, Auriane Técourt, Cherry Wu, Catherine Rhodes, Jess Bland, Peter Glenday, and others. We additionally thank all of the participants who have played Intelligence Rising for providing us with the experiences that we drew upon for this study.

# Funding Statement

Funding for the research conducted in this paper was initially provided by the Long Term Future Fund, and subsequent funding has been provided by a donation from Ramana Kumar.

# Author contributions

Author contributions are presented according to the International Council for Medical Journal Editors criteria for authorship.

Contribution to conception or design of the research: RG

Contribution to acquisition of the research: SA, RG, JF, AS

Contributed to analysis or interpretation of data for the research: RG, SA, JF, AS

Drafted the research or revisited the research critically for important intellectual content: RG, SA, JF, AS

Provided final approval of the research to be published: RG, SA, JF, AS

Agrees to be accountable for all aspects of the research in ensuring that questions related to the accuracy or integrity of any part of the research are appropriately investigated and resolved: RG, SA, JF, AS

# Conflicts of interest

Facilitators were paid a fixed rate for the games that involved paying clients; however, roughly half of the games covered were unpaid, and of the paid games, facilitators often chose to facilitate these games pro bono. Intelligence Rising's assets and intellectual property were transferred in 2024 from individuals contributors to Technology Strategy Roleplay, a U.K. charity (Charity number: 1200928) that was established to manage Intelligence Rising and other efforts by the Intelligence Rising team to create games for exploring AI futures and raising awareness about the risks of AI.

## Appendix A: List of Games Considered in Analysis

| Date | Facilitator(s) | Client Type | Client location | Paid? |
|---|---|---|---|---|
| 09/2020 | Shahar | Student group | Singapore | No |
| 09/2020 | Shahar | NGO | USA | No |
| 12/2020 | Shahar | Other | UK | No |
| 12/2020 | Shahar | Insurance company | France | No |
| 02/2021 | Shahar | Academic group | UK | No |
| 05/2021 | Shahar, James | University course | UK | Yes |
| 05/2021 | Shahar, James | University course | UK | Yes |
| 05/2021 | Shahar, Ross | AI company | US | Yes |
| 09/2021 | Shahar | AI company | UK | No |
| 10/2021 | Shahar, James | Conference | UK | No |
| 01/2022 | Shahar, James | University course | UK | Yes |
| 05/2022 | Shahar | Academic group | UK | No |
| 06/2022 | Shahar | Student group | UK | No |
| 07/2022 | Shahar, James | University course | UK | Yes |
| 07/2022 | Shahar, James | University course | UK | Yes |
| 08/2022 | Shahar | Student group | US | Yes |
| 09/2022 | Shahar | Non-academic training | Belgium | Yes |
| 12/2022 | Shahar | NGO | Remote | No |
| 05/2023 | Shahar, James, Ross | AI company | UK | No |
| 10/2023 | Shahar | Insurance company | UK | Yes |
| 7/11/23 | Ross | Think tank | US | Yes |
| 7/10/23 | Ross | Think tank | US | Yes |
| 4/21/23 | Ross, James | Conference | Sweden | No |
| 4/22/23 | Ross, James | Conference | Sweden | No |
| 3/3/23 | Ross | Conference | US | Yes |
| 10/22/22 | Ross | Student group | Remote | Yes |
| 4/29/22 | Ross | AI company | Remote | Yes |
| 4/17/22 | Ross | Conference | UK | No |
| 3/20/22 | Ross | Student group | Remote | No |
| 3/1/22 | Ross | Non-academic training | Remote | No |
| 10/21/21 | Ross | Other | Remote | No |
| 8/26/21 | Ross | Academic group | Remote | No |
| 2/27/21 | Ross | Non-academic training | Remote | No |
| 08/2022 | James | Other | UK | No |
| 08/2022 | James | Non-academic training | UK | Yes |
| 08/2023 | James | NGO | US | Yes |
| 01/2023 | James | University course | UK | Yes |
| 02/2023 | James | Student group | UK | Yes |
| 12/2023 | James | NGO | Belgium | Yes |
| 01/2024 | James | University course | UK | Yes |
| 03/2024 | James | Non-academic training | Belgium | Yes |
| 05/2024 | James | NGO | UK | No |
| 07/2024 | James | Non-academic training | UK | Yes |